**Strain-induced topological phase transition in two-dimensional platinum ditelluride**

Jiesen Li,[1] Wanxing Lin,[2,*] and D. X. Yao[2,†]

[1] School of Environment and Chemical Engineering, Foshan University, Foshan, P. R. China

[2] State Key Laboratory of Optoelectronic Materials and Technologies, School of Physics, Sun Yat-sen University, Guangzhou, P. R. China

**Abstract**

Topological phase transition is a hot topic in condensed matter physics and computational material science. Here, we investigate the electronic structure and phonon dispersion of the two-dimensional (2D) platinum ditelluride ($PtTe_2$) using the density functional theory. It is found that the $PtTe_2$ monolayer is a trivial insulator with an indirect band gap of 0.347eV. Based on parity analysis, the biaxial tensile strain can drive the topological phase transition. As the strain reaches 19.3%, $PtTe_2$ undergoes a topological phase transition, which changes from a trivial band insulator to a topological insulator with $Z_2 = 1$. Unlike conventional honeycomb 2D materials with topological phase transition, which gap closes at K points, the strained $PtTe_2$ monolayer becomes gapless at M points under critical biaxial strain. The band inversion leads the switch of the parities near the Fermi level, which gives rise to the topological phase transition. The novel monolayer $PtTe_2$ has a potential application in the field of micro-electronics.

---

\* Corresponding author. E-mail: linwanhsing@foxmail.com (Wanxing Lin)

† Corresponding author. E-mail: yaodaox@mail.sysu.edu.cn (D. X. Yao)

**Introduction**

Two-dimensional (2D) materials have attracted great attention because of their potential applications in micro-electronics, which arises from their particular electronic properties such as flexibility, high carrier mobility, and low effective mass. Among those 2D materials under intensive studies, transition metal dichalcogenides (TMDs), with chemical formal $TX_2$ (T is transition metal, and X=S, Se, Te), exhibits a wide range of properties, from insulator to superconductor, depending on the composition [1,2,3,4]. The study of $PtX_2$ (X=S, Se, Te) dates back to the 1950s. The thermal expansion was measured in 1959[5], the electronic structures were studied in 1986 [6], and their multilayers were later used for electrochemical analysis [7]. In particular, $PtTe_2$ was found to be a type-II Dirac semimetal, which was confirmed by angle-resolved photoemission spectroscopy (ARPES) [8,9,10,11,12,13]. Afterwords, the multilayers of $PtTe_2$ was synthesized or exfoliated, the surface was found to be stable [14,15,16]. Layered $PtTe_2$ can catalyze oxygen reduction reaction with performance comparable to platinum on carbonwith significantly lower toxicity [17]; on the other hand, defects on the Te sites can promote the hydrogen evolution catalysis [18]. Besides, transition metal dopped monolayer $PtTe_2$ may show ferromagnetism [19]. Monolayer $PtTe_2$ may also be used as electronic tattoos in medical devices [20]. When intercalated with potassium, bilayer $PtTe_2$ shows enhanced superconductivity [21]. The electronic structure of $PtTe_2$ single crystals was found to be thickness-dependent [22]. The electronic structures of $PtTe_2$ as well as its elastic and optical properties were also studied, and it was found that strain can induce a semiconductor-to-metal transistion [22,23,24]. The strained $PtTe_2$ monolayer also exhibits enhanced thermoelectric properties[25]. However, the topological properties of $PtTe_2$ under strain have not been studied.

In this work, we investigate the electronic structure of $PtTe_2$, and its strain-dependence from the topological point of view. Since $PtTe_2$ has inversion symmetry, with the Pt atom being the inversion center, we can determine the symmetry of the wavefunctions under various strains. $Z_2$ is found to change from 0 to 1 under uniform biaxial tensile strain, which means $PtTe_2$ changes from a trivial insulator to a topological insulator.

**Model and calculation details**

We model the 2D system by inserting a vacuum at least 15Å thick, so as to avoid the coupling between periodic images. The calculations are based on the density functional theory (DFT), as implemented in the Vienna *Ab initio* Simulation Package (VASP) code [26] and RESCU code [27]. Plane augmented wave (PAW) is used as a basis, and Perdew-Burke-Ernzerh (PBE) is treated as the exchange-correlation [28]. The Brillouin zone is sampled by $20 \times 20 \times 1$ Monhost-Pack grid. For phonon calculation, we use $4 \times 4 \times 1$ supercell in the finite displacement scheme, and double zeta numerical atomic orbital. The Spin-orbit coupling (SOC) is included in the electronic structure calculations.

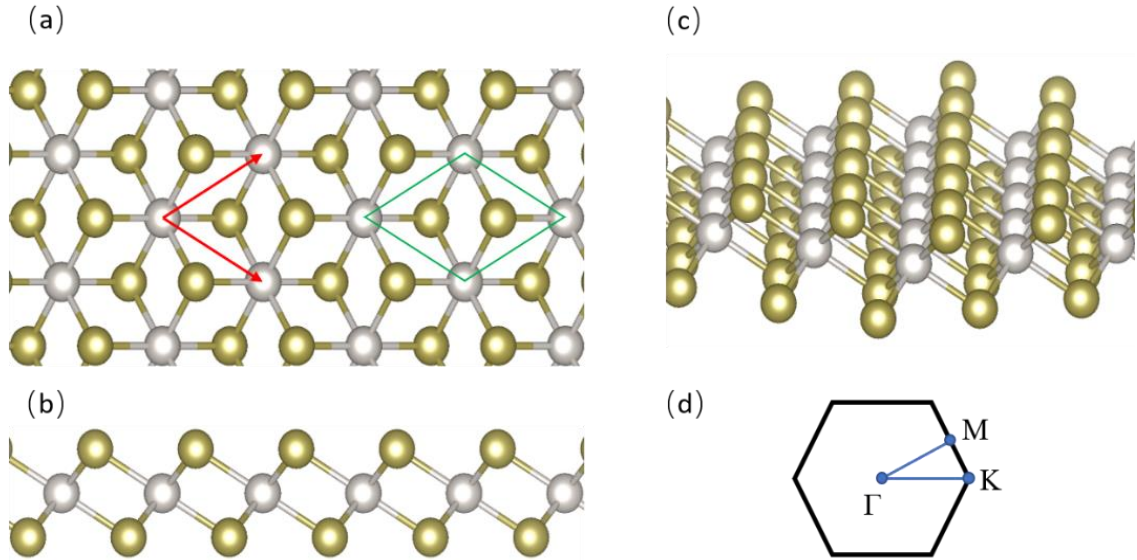

Figure 1. The (a) top view, (b) side view, and (c) bird's eye view of PtTe$_2$ 2D structure. Grey and dark yellow spheres correspond to Pt and Te atoms, respectively. Red arrows denote basis vectors, and the green rhombus denotes the unit cell. (d) The first Brillouin zone and the high symmetry points.

**Structure and stability**

The crystal structure of the PtTe$_2$ monolayer is shown in Figure 1. PtTe$_2$ monolayer has a inversion symmetry belonging to $D_{3d}$ point group, and the Pt atom is the center of inversion. The lattice constant is 4.02Å, the Te-Pt bond length is 2.70Å. The Te-Pt-Te bond angle is 84.12º, the Pt-Te-Pt bond angle is 95.88º, and the buckling distance is 2.79Å. In order to verify the stability of the monolayer, we calculated the phonon dispersion, as shown in Figure 2. No imaginary frequency is found and thus is dynamically stable. These results imply the possibility of isolating atomically thin monolayer of PtTe$_2$.

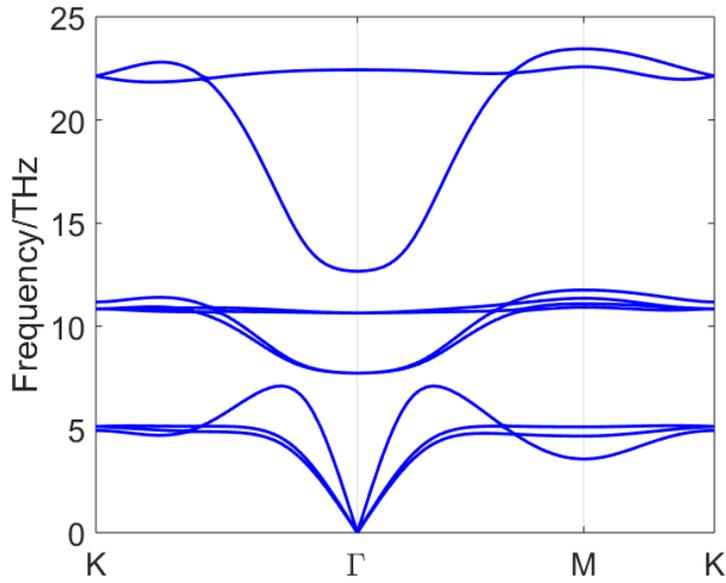

Figure 2. Phonon dispersion of PtTe₂ 2D monolayer.

**Electronic structures and the topological phase transition**

From the projected density of states (DOS) (Figure 3), the states near the Fermi energy are dominantly composed of d orbitals of Pt atoms and p orbitals of Te atoms. The s orbital character of both elements appears far away from the Fermi energy; thus, they play an insignificant role in the electronic structures. Specifically, the inner bands have a strong s character of Te atom, while the ones with high energy have a strong s character of Pt atom. The bands near the Fermi level are mainly composed of d orbital of Pt and p orbital of Te. From the band structure, unstrained PtTe₂ is a semiconductor with an indirect band gap of 0.347eV. Since the PBE pure functional tends to underestimate the band gaps of semiconductors, the actual band gap should be wider than our results. The valance band maximum (VBM) is located in Γ point, while the conducting band minimum (CBM) is located somewhere between Γ and M point.

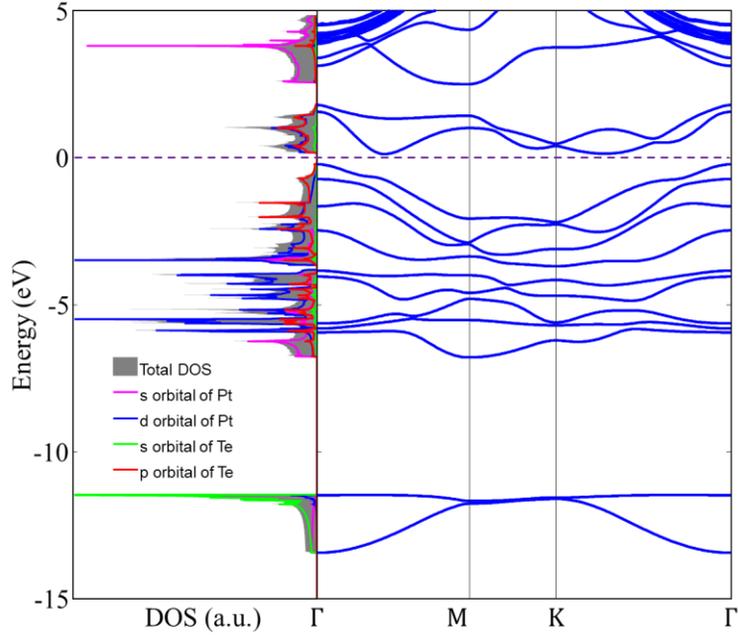

Figure 3. Projected DOS (left panel) and band structure (right panel) of unstrained PtTe$_2$ monolayer.

Table 1

| Band Label | No strain | | 10% strain | | 19.3% strain | | 20% strain | |
|---|---|---|---|---|---|---|---|---|
| | Γ | M | Γ | M | Γ | M | Γ | M |
| 1 | + | + | + | + | + | + | + | + |
| 2 | + | + | + | + | + | + | + | − |
| 3 | + | + | + | + | + | − | + | + |
| 4 | + | − | + | − | + | + | + | + |
| 5 | + | + | + | + | + | − | + | − |
| 6 | + | + | + | + | − | + | − | + |
| 7 | − | + | − | + | − | + | − | + |
| 8 | − | − | − | − | + | + | + | + |
| 9 | − | − | − | − | − | + | − | + |
| 10* | + | + | + | + | + | − | + | − |
| $\prod_{occ}(-1)^{\nu_i}$ | − | − | − | − | − | + | − | + |
| $Z_2$ | 0 | | 0 | | 1 | | 1 | |
| * This band is above the Fermi level. | | | | | | | | |

The band structure gradually changes as the biaxial strain increases. As the biaxial strain increases, the gap increases and reaches a maximum at 4%, and then decreases and eventually closes at 19.3%, as shown in Figure 4. During this process, the VBM deviates from the $\Gamma$ point, and a local maximum appears on both $\Gamma - M$ line and $\Gamma - K$ line. These two local maxima have different energies, therefore, there are two band gaps, denoted as $E_{g1}$ and $E_{g2}$, and $E_{g1} > E_{g2}$; while $E_{g1}$ is direct, $E_{g2}$ is indirect. Meanwhile, the CBM gradually moves to M point, then the valance band and the conduction band eventually come into contact at the critical strain of 19.3%. Then the gap reopens as the biaxial strain continues to increase.

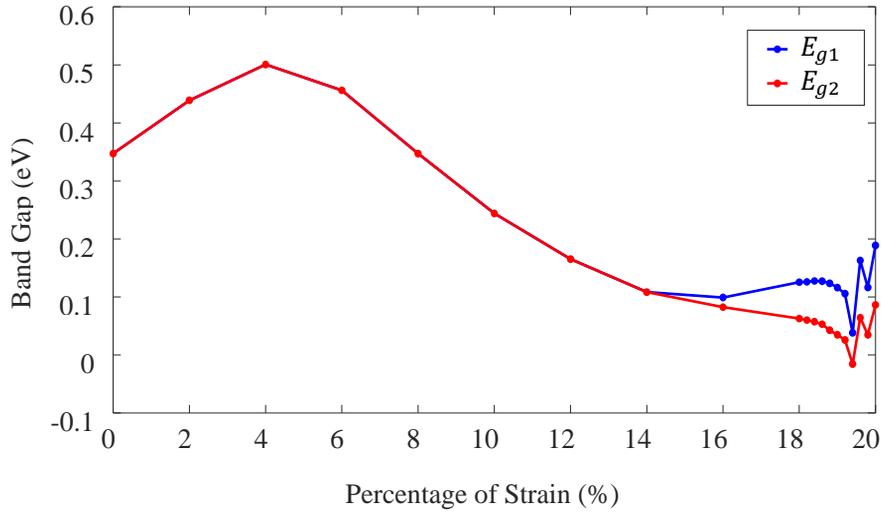

Figure 4. The strain-dependence of band gap. The blue and red lines are $E_{g1}$ and $E_{g2}$, respectively.

Because of the inversion symmetry, we can use $Z_2$ to classify the topological states under various strains [29]. To investigate the topologies of the band evolution, we calculated the parities of the wavefunctions at the Time-Reversal Invariant Momenta (TRIMs), and thus calculated the $Z_2$. The system has four TRIMs in the Brillouin zone — three M's and a $\Gamma$. The parity at any particular TRIM is determined by the parities of all the occupied states, that is

$$(-1)^\nu = \prod_{occupied} (-1)^{\nu_i}.$$

The $Z_2$ of the system is the product of the parities at all the TRIMs. The parities band structure under various biaxial strains are also labeled in Figure 4. All parities of the wavefunctions at the four TRIMs are listed in Table 1. Because of crystal symmetry, the three M points have the same parity. As the strain gradually increases, the parities on the four TRIMs are unchanged, regardless of the minor shifts of the bands. At the critical strain of 19.3%, when the gap closes and reopens, the parities at the M points change; Particularly, the band inversion leads to the interchange of the parities between the states just above and below the Fermi level, as shown in Figure 5(c-d). This inversion gives rise to transition from atrival insulator to a topological insulator, in that the $Z_2$ changes from 0 to 1. The evolution of the band structure near the Fermi level before and after the transition can be depicted schematically in Figure 6. The SOC does not dramatically change the band structure before the gap reopens. In Figure 6(c), after gap closing, the band structure without SOC is shown in a dashed black line, and it is gapless; however, SOC opens a gap, as the solid red line shows. In other words, SOC plays an important role in the band inversion and the topological phase transition.

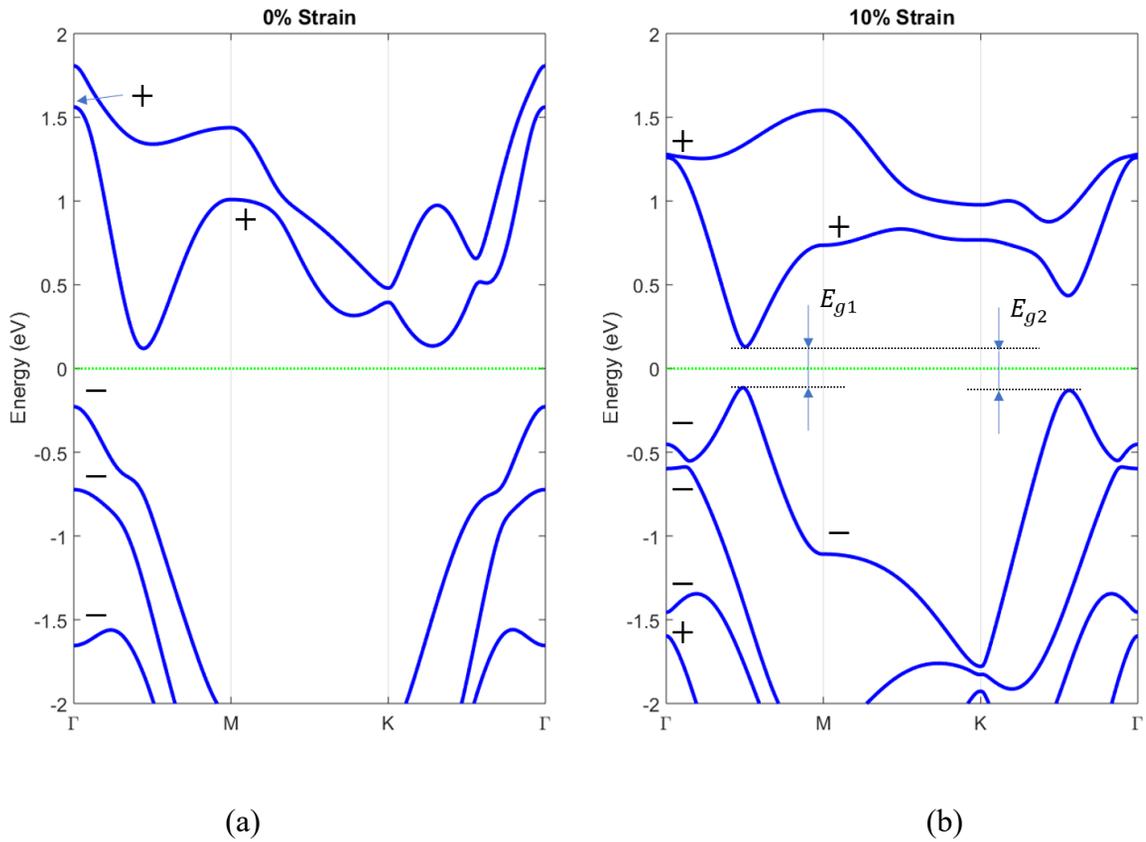

(a)  (b)

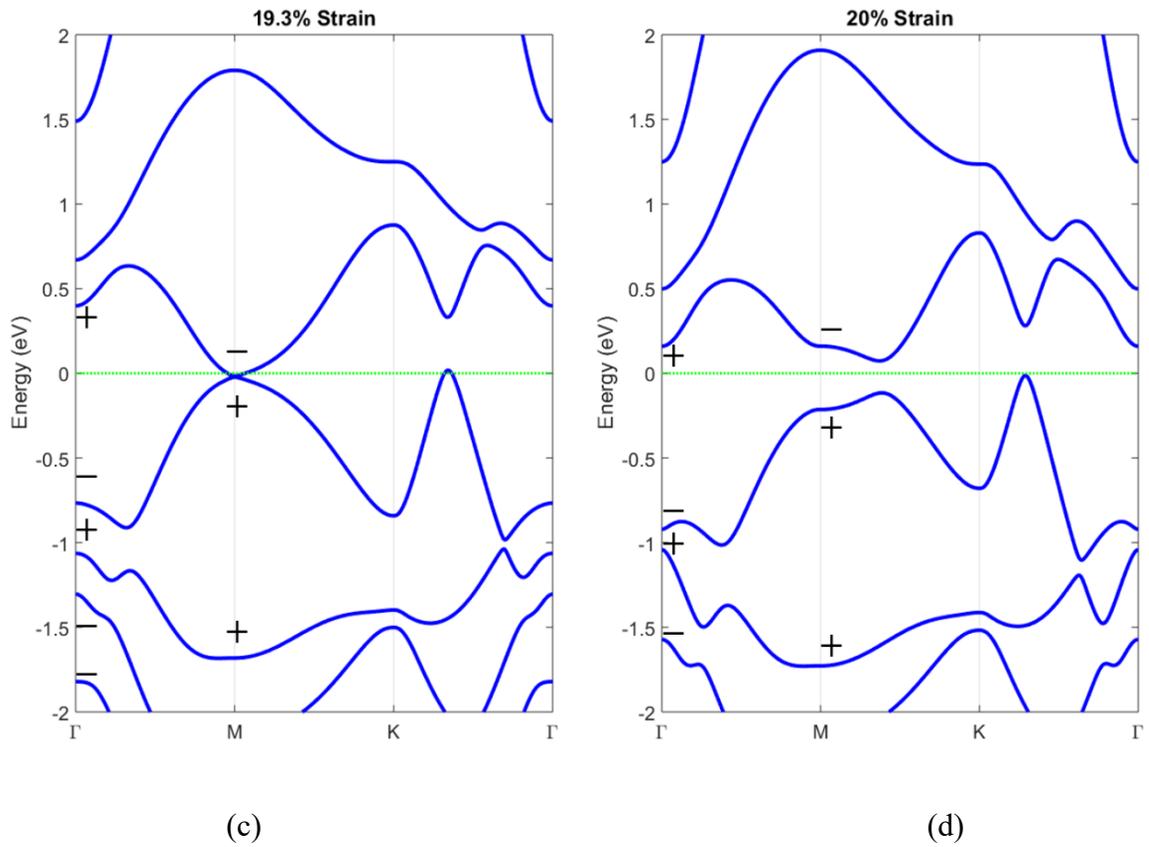

Figure 5. Band structures under (a) 0%, (b) 10% (b), (c) 19.4%, and (d) 20% biaxial strains, respectively. Parities of wavefunctions at TRIMs are labeled as "+" or "−". Definition of $E_{g1}$ and $E_{g2}$ are shown in (b).

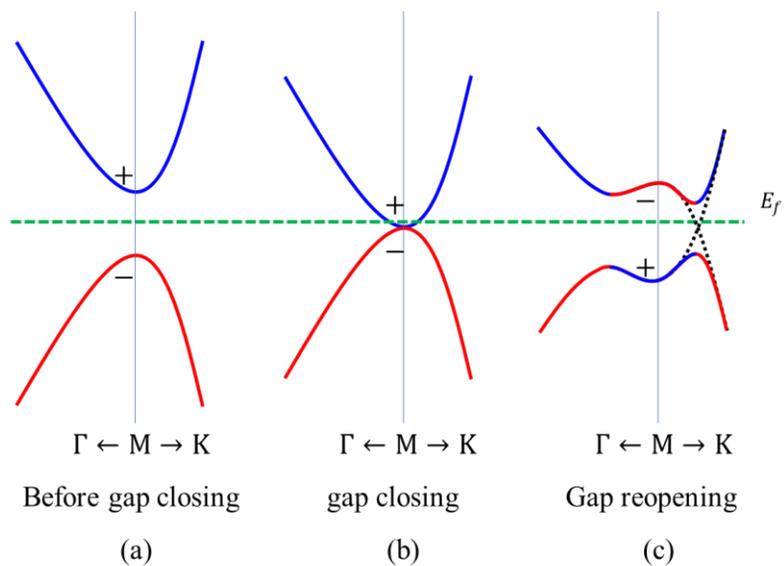

Figure 6. Schematic drawing of band evolution around M point. (a) Band structure of trivial insulator. (b) Band structure at the critical points. (c) Band structure of topological insulator. The dashed black

lines represent bands without SOC, and solid red line represents bands with SOC.

**Conclusion**

In this paper, we have studied the electronic structures of atomically thin PtTe$_2$ monolayer from the topological perspective. PtTe$_2$ is an insulator with an indirect band gap of 0.347eV, and the d orbital of Pt and the p orbital of Te are the main components of the states near the Fermi level. From the symmetry analysis of the four TRIMs, we find $Z_2 = 0$, meaning it is a trivial band insulator. When the biaxial tensile strain is exerted on the monolayer, the band gap increases initially, then decreases and eventually closes at 19.3% strain at M points. Before gap closing, the system remains a trivial insulator with $Z_2 = 0$. As the strain exceeds 19.3%, the gap reopens. Band inversion leads to the changes of the parities near the Fermi level at M points, and the system changes from a trivial insulator into a topological insulator with $Z_2 = 1$. Further analysis reveals that SOC plays an essential role in this transition. Our results may serve as a reference for the electronic application of PtTe$_2$ monolayer.

**Acknowledgments**

W. L. and D. X. Y. are supported by NKRDPC-2017YFA0206203, NKRDPC-2018YFA0306001, NSFC-11974432, GBABRF-2019A1515011337, and Leading Talent Program of Guangdong Special Projects.